\RequirePackage{fix-cm}
\documentclass[smallextended]{svjour3}       % onecolumn (second format)
\smartqed  % flush right qed marks, e.g. at end of proof
\usepackage{graphicx}
\usepackage{amssymb, amsmath, mathrsfs, amsfonts, multirow, color}

\usepackage[colorlinks=true, pdfstartview=FitV, linkcolor=blue, citecolor=blue, urlcolor=blue]{hyperref} 

\newcommand{\E}{{\mathbf{E}}}

\journalname{Advances in Austrian Economics}
\begin{document}

\title{Game Mining:%\thanks{Grants or other notes
}
\subtitle{How to Make Money from those about to Play a Game}


\author{James W. Bono         \and
       David H. Wolpert%etc.
}


\institute{James Bono \at
					Microsoft,	Redmond WA\\
              \email{james.bono@microsoft.com}           %  \\
           \and
%           David H. Wolpert \at
%Information Sciences Group\\
%Los Alamos National Laboratory, NM\\
%\email{david.h.wolpert@lanl.gov}\\
%and \\
%Santa Fe Institute, NM \\
%\email{dhw@santafe.edu}
%}
           David H. Wolpert \at
%Information Sciences Group\\
%Los Alamos National Laboratory, NM\\
%\email{david.h.wolpert@lanl.gov}\\
%and \\
Santa Fe Institute, NM, \\
%and \\
Complexity Science Hub, Vienna, \\
%and \\
International Center for Theoretical Physics, Trieste, \\
%and \\
Arizona State University, Tempe AZ, \\
\email{david.h.wolpert@gmail.com}
%\web{davidwolpert.weebly.com}
}

%\date{Received: 8/15/2011 / Accepted: date}
\date{Received: 8/15/2011 / Accepted: 2013}

\maketitle

\begin{abstract}
It is known that a player in a noncooperative game can benefit by publicly
restricting his possible moves before play begins. We show that, more generally, a player may benefit by publicly committing to pay an external party an amount that is contingent on the game's outcome. We explore what happens when external parties -- who we call ``game miners'' -- discover this fact and seek to profit from it by entering an outcome-contingent contract with the players. We analyze various structured bargaining games between miners and players for determining such an outcome-contingent contract. These bargaining games include playing the players against one another, as well as allowing the players to pay the miner(s) for exclusivity and first-mover advantage. We establish restrictions on the strategic settings in which a game miner can profit and bounds on the game miner's profit. We also find that game miners can lead to both efficient and inefficient equilibria.
\end{abstract}

\section{Introduction}
\subsection{How to Mine a Game}
That players can benefit in games by entering contracts that distort
payoff functions is well-documented in the economic literature [see
\cite{Schelling56,Sobel81,Vickers85}]. In this paper we focus on a
special case of this phenomenon: A player $i$ may benefit by publicly
committing to pay an external party an amount that is contingent on
the game's outcome. In this paper we explore what happens when
external parties discover this fact and seek to profit from it.

To ground the discussion, we present an example.

\begin{example}

There are two cell-phone manufacturers, Anonymous (A) and Brandname (B). They must simultaneously decide how many cell phones to produce. Each firm has two options, high output (H) or low output (L). Anonymous, like its name suggests, is not well known. Therefore, no matter what level of output Brandname produces, Anonymous prefers to produce high output to gain brand recognition. On the other hand, Brandname's choice of output does depend on Anonymous's choice. If Anonymous produces low output, then Brandname prefers to keep prices high by also producing low output. However, if Anonymous produces high output, then Brandname prefers to safeguard its recognized name by also producing high output. The moves and payoffs (in millions of dollars) are summarized by the following matrix.
$$\begin{array}{c c|c|c|}
\multicolumn{2}{c}{} & \multicolumn{2}{c}{B}\\
\multicolumn{2}{c}{} & \multicolumn{1}{c}{{\mbox{H}}} &
\multicolumn{1}{c}{{\mbox{L}}} \\
\cline{3-4}
\multirow{2}{*}{A} & {\mbox{H}} & 1,.5 & 2,0 \\
\cline{3-4}
& {\mbox{L}} & 0,0 & 1,1 \\
\cline{3-4}
\end{array}$$
The NE is $(H, H)$, and payoffs are $(1,.5)$.

A firm called Game Mining Inc. (G) watches this, and just before Anonymous and Brandname declare their output decisions, $G$ offers Anonymous the following contract, making sure Brandname sees them do this: ``Pay us \$1.5 million right now. Then we will pay you back a certain amount after you and Brandname make your decisions. The following matrix shows how much we will pay you, in millions of dollars, for the four possible joint decisions by you and Brandname:"
\begin{equation*}
\left[\begin{array}{ c c }
0 & 1.01 \\
1.5 & 2.0
\end{array}\right].
\end{equation*}
Note that this matrix does not include the \$1.5 million up front payment.
So if Anonymous accepts the contract, then the final payoff matrix for the game (including the up front payment) becomes
$$\begin{array}{c c|c|c|}
\multicolumn{2}{c}{} & \multicolumn{2}{c}{B}\\
\multicolumn{2}{c}{} & \multicolumn{1}{c}{{\mbox{H}}} &
\multicolumn{1}{c}{{\mbox{L}}} \\
\cline{3-4}
\multirow{2}{*}{A} & {\mbox{H}} & -.5,.5 & 1.51,0 \\
\cline{3-4}
& {\mbox{L}} & 0,0 & 1.5,1 \\
\cline{3-4}
\end{array}$$

In this modified game, the unique NE is very close to
$(2/3, L)$, where $2/3$ is the probability that $A$ chooses $H$. For Anonymous, this results in an average payoff of approximately \$1.5 million. This is a \$500,000 improvement over its payoff without the contract. On average, $G$ makes approximately \$160,000 for their trouble. Brandname, on average, makes approximately \$333,333. This is a \$166,666 decrease in Brandname's payoff compared to the situation where Anonymous and $G$ do not have a contract.
\label{ex:2}
\end{example}

Note that the Coaseian outcome of the game without $G$ would be for Anonymous to pay at least \$500,001 to Brandname for the outcome $(H,L)$ [see \cite{Coase60}]. So $G$ is not merely facilitating the Coaseian outcome. Note also that the outcome in example \ref{ex:2} cannot be achieved by a commitment by either player to play or not play either one of their actions.  In general, the presence of $G$ creates an entirely new strategic setting.  However, in this simplest of game mining scenarios, a commitment by $A$ to play the mixed strategy $(2/3,1/3)$ would achieve the same equilibrium as the game miner's outcome-contingent contract. This is not the case for the other game mining scenarios discussed in this paper.

In example \ref{ex:2} $G$ makes considerable profit from recognizing that Anonymous benefits  from an output-contingent contract. However, Anonymous also benefits from this contract.  So game mining can be seen as a natural result of profit-seeking behavior, rather than as the result of economic planning and regulation. It requires no mechanism other than enforceable contracts between players and game miners. 

Since profit-seeking behavior creates the game mining setting in example \ref{ex:2}, it is natural to think about what other types of game mining settings can arise this way.  By studying various structured bargaining games between game miners and players, we find a range of new phenomena. For example, in some cases, it may be that the miner can create a prisoner's dilemma scenario between the players by offering contracts that each player has an incentive to accept regardless of whether their opponent accepts. If both players accept, the outcome is that they are both worse off than if neither accepts, and the game miner profits considerably.  There are also important timing issues in game mining. When players sequentially sign contracts with game miners, there can be a significant first-mover advantage to the first-signing player. This provides the game miner with yet another opportunity for profit; they can charge the players to move first.  Finally, game mining has complicated effects on efficiency. The effects depend on the underlying game as well as the bargaining structure between players and the game miner. In this paper, we explore situations where the presence of a game miner can both increase and decrease equilibrium efficiency.

%It is also worth noting that in many mining scenarios, \textit{mixed contracts} are needed to
%guarantee the existence of equilibria. With such contracts, players
%have uncertainty about their opponent's payoffs. However, unlike in
%Bayesian Nash equilibrium, in game mining this uncertainty is resolved before the
%game is played. Its only role is to make the players indifferent among
%their own contracts. That is, a player will say, ``Given that you will
%select a contract (and therefore an ultimate payoff function)
%according to that probability distribution, I am indifferent among the
%following contracts (and therefore ultimate payoff functions). So I
%will randomize over them with this probability distribution, which in
%turn makes you indifferent among the contracts in the support of your
%probability distribution.''

\subsection{Related Literature}

The ideas underlying the game mining concept are implicit in a large body of economic literature. As an illustration, in the model presented in \cite{Jackson05} (JW), every player specifies outcome-contingent side-payments that they will make after a non-cooperative
strategic form game is played and the payoffs are resolved. These side-payments are binding contracts, so the players are \emph{ex ante} determining their preferences over the game's outcomes. In this regard the game that the players actually play is endogenously determined. JW examine whether a mechanism that allows players to make such outcome-contingent side-payments generally results in efficient outcomes and conclude that it does not. 

The simplest game mining scenario, e.g., example \ref{ex:2}, can be viewed as a variant of JW. In this variant, the only outcome-contingent side-payments are between the players and the game miners and the game miners would be indifferent over outcomes of the game if not for the fact that they will be receiving side-payments dependent on those outcomes. Furthermore, the game miners play no part in the game between the players other than to accept contracts for outcome-contingent side-payments and make the contracts public. 

In contrast to JW, we do not assume that a social planner installs a mechanism for players to make side-payments. Instead, we look at a game without formal mechanisms and ask whether external parties will create contracts for outcome-contingent side-payments in pursuit of profit. In addition, we relax the assumption in JW that all side-payments are nonnegative. That is, we allow game miners to pay players for certain outcomes. This will be important when examining optimal contracts as well as the extent to which a monopolist game miner can extract profits from players. Hence, game mining can be seen as the version of JW's setting that arises as the natural result of profit-seeking in the absence of JW-type regulation. We also consider how outcomes change when the game miner can enter contracts with both players, when mining contracts are offered to players in sequence, and when we change the underlying bargaining structure between the players and the game miner.  These considerations do not arise in JW's setting.

In another related paper, \cite{Renou09} analyzes what happens when players are able to embed the original game in a new two-stage game. In the second stage of that new game the players play the original game. However before they do so, in the first stage, the players each simultaneously commit not to play some subset of their possible moves in that game in the second stage. These \textit{commitment games} can be seen as the result of placing three restrictions of the JW setting: (1) player $i$'s side-payments are only contingent on $i$'s action (rather than on the full profile of actions), (2) the side-payments are made to external players and (3) the side-payments are effectively infinite. Renou's restricted setting means that there are fewer commitments available but that there are also fewer deviations available. This means the set of equilibria sustainable in Renou's restricted setting can differ significantly from the set sustainable in the less restricted JW setting. Renou's setting and analysis are too restricted to cover the full game mining scenario. This is because there are many circumstances in which both the player and the game miner prefer to make contracts that are fully outcome-contingent and that have non-infinite side-payments. One example can be found in example \ref{ex:2} above, where the player and game miner find it optimal to agree on a contract that results in a unique equilibrium in which the player uses a mixed strategy with full support (and therefore does not make any commitment in the first stage of Renou's two-stage game).

In \cite{Garcia06} the types of commitments are similar to those studied in \cite{Renou09}. The main difference is that the former studies a repeated game in which commitments to play mixed strategies are possible. They show that when such commitments are available, standard folk theorem results can be obtained with milder assumptions on the stage game.  A key difference between our work and that of \cite{Garcia06} is that our settings are all single-shot. Still, it should be possible to make commitments to play mixed strategies in a single-shot game.  In the simplest game mining scenario, the set of equilibria is the same as what can be sustained by commitments to play mixed strategies. However, when mixed strategies do arise in our framework, they do so as the result of contracts that produce games with unique mixed NE rather than being required by the commitments themselves. And beyond the simplest game mining scenario, the set of equilibria under commitments to play mixed strategy differ from game mining equilibria. Like \cite{Garcia06}, the work of \cite{Kalai10} also obtains a folk theorem for commitments. They study a setting in which players can make commitments that are conditional on the commitments of their opponents. For two-player strategic games, they show that all feasible and individually rational payoffs can be obtained in the equilibrium set of a game with conditional commitments.  Such conditional commitments are not explored in the game mining framework but remain of interest for future work. 

Another well-studied aspect of commitment in games is the role of timing. Papers such as \cite{Hamilton90}, \cite{vanDamme96} and \cite{Romano05} concern endogenous timing and Stackelberg-like commitments. \cite{Wolpert11,Wolpert12} introduce the idea that players might also commit to using certain personas before the start of a game. These papers analyze the possibility that experimentally observed non-rationality is in fact rational, because by committing to play the game with a non-rational ``persona'', a player may increase her ultimate payoff. This persona has the same effect as a side-payment or commitment, as it is reflected in a temporary change to the player's utility function. The \textit{persona games} model has been successful in explaining non-rational behavior in non-repeated traveler's dilemma and even in versions of the non-repeated prisoner's dilemma. Timing is also studied in the current paper. Here, the focus is on how the sequence of contracts between the game miner and players determines the outcome.  As is the case in persona games, there is often a first-mover advantage in game mining even when players have dominant strategies. 

Finally, there is a subset of the principal-agent literature concerning delegation games that is closely related to game mining. In these models, the principal is able to contract with an agent that will engage in a game with the principal's opponent (or agent of the principal's opponent). One concern of this literature is detailing the optimal contract for a principal [see \cite{Vickers85,Fershtman85,Sklivas87}]. Another concern is whether a mechanism that allows specific types of contracts can lead to Pareto efficiency [see \cite{Kalai91,Katz91}]. Game mining is closely related to a previously unexplored aspect of principal-agent scenarios: the degree to which the agents can profit from delegation contracts.

The important difference between the work here and the related literature is that we concentrate on the potential role of a third party, seeking to make profits by interacting with the players, rather than focus on situations where all relevant decisions are by the original players alone.

\subsection{Overview}

We start by introducing the game mining model and notation in section \ref{sec:notation}. Then, in section \ref{sec:maxmine} we assume one player interacts with the game miner and conduct a foundational analysis of the types of contracts and outcomes they can achieve. The properties derived in section \ref{sec:maxmine} serve as the foundation for more in-depth analysis that appears later in the paper. For example, here we derive bounds on the aggregate payoffs that the game
miner and player can earn together. We show that they can select a contract to
divide these payoffs in any way between them. We also show that
outcome-contingent contracts cannot be profitable for both the game
miner and contracting player if the other player has a strictly dominant strategy.

In section \ref{sec:structures} we consider various market structures, i.e.,
various structured bargaining games involving the players of
the underlying noncooperative game and an external party
that tries to mine that underlying game. We begin in section \ref{subsec:onecontract} with a game based on the
assumption that players offer contracts to the game miner and the game
miner must choose either one of the offered contracts or neither
contract. We show that the game miner can profit by more than the
maximal payoff to either player in the game without contracts. This is
because players can suffer a loss if their opponent outbids them for
the right to contract with the game miner. We derive an upper bound on the game miner's equilibrium profit. We also characterize the set of equilibria for a simple example and show that the opportunity to make contracts with the game miner can make both players worse off.  That is, a Pareto efficient equilibrium of the underlying game is not an equilibrium when game mining is possible. Note that JW find a similar result using their contingent transfers mechanism.

In section \ref{subsec:bothcontracts}, we allow the game miner
to accept both offers if she so chooses. This reduces the game miner's
bargaining power, and we find that the game miner can always do at
least as well by restricting herself to accept only one contract. 

In section \ref{subsec:sequential}, we discuss the role of timing and first-mover advantage, establishing
that the players may be willing to pay for the right to contract first
with the game miner. Therefore, in contrast to the simultaneous contract case in section \ref{subsec:bothcontracts}, $G$ can engage players in a bidding war even when she can accept both contracts. Furthermore, this can happen when one or both of the players have a strictly dominant
strategy. Two examples illustrate this and further analyze the difference between game mining and Renou-type commitments. 

In section \ref{subsec:Gmakes} we analyze the case where the miner has the
bargaining power, i.e., $G$ is the one making the offers. We show that
this allows the game miner to ``play the players against one another''
and thereby increase her profit. We also derive an upper bound on this
profit. An interesting feature of this scenario is that the game miner moves the players from an inefficient equilibrium to an efficient one. However, as a profit seeker, the game miner is able to capture \emph{more} profit than the efficiency gain.

%In section five we look at perfect competition and duopoly miner
%market structures. We develop the notion of best contract response
%correspondences and mixed contracts to discuss the existence of
%equilibria when players simultaneously choose contracts. We also
%detail the way in which a duopolist game miner's profits depend on the game that arises as a result of contracts.

In section \ref{sec:discussion} we briefly discuss several new research areas opened by
game mining. These areas include games of more than two players, and 
risk aversion on the part of the game miner. We briefly consider
unstructured bargaining among the players and the
miner to determine the contract. We also touch on an ``inverted" version of this topic, where
the underlying game is itself unstructured, while the miner(s) negotiate with
the player(s) via structured bargaining to determine a contract for that underlying game.
We end by discussing the idea that one player might sign a contract that obligates him to pay the other player outcome-contingent amounts. This obligation may actually help the payer and hurt the payee.

\section{Notation}
\label{sec:notation}
We study a two-player, one-stage simultaneous-move game of complete
information. However, we relax the usual assumption that the two players
cannot make outcome-contingent contracts (or simply contracts) with
players external to the game.

Specify the two-player \emph{pre-contract} game as
$$\Gamma\equiv \{A,B\},\{X_A,X_B\}, \{U_A,U_B\}),$$
where $X_i\in \{A,B\}$ is finite and $U_i$ is an $|X_A|$-by-$|X_B|$ matrix for which the $(m,n)$ entry gives the payoff
to $i$ when $A$ chooses his $m$'th pure strategy and $B$ chooses his
$n$'th pure strategy. Player $i$'s set of mixed strategies is
$\Delta_i$, $i=A,B$, and the set of mixed strategy profiles is
$\Delta\equiv\Delta_A\times\Delta_B$. We write all of $i$'s pure and mixed
strategies as $|X_i|$-by-one vectors $\sigma_i$ for which the $m$'th
entry gives the probability that $\sigma_i$ assigns to playing $i$'s
$m$'th pure strategy. Therefore, player $i$'s expected
payoff from $\sigma=(\sigma_A,\sigma_B)$ as $$
\E_\sigma(U_i) = \sigma_A^T U_i\sigma_B
$$
where superscript$-T$ indicates matrix transpose.

Player $i$'s best response correspondence is given by $R_i^\Gamma
(\cdot):\Delta_j\rightarrow 2^{\Delta_i}$, so that
\begin{equation}\label{defBR}R_i^\Gamma(\sigma_j)\equiv\{\sigma_i\in \Delta_i:\quad \sigma_A^T U_i\sigma_B\geq \sigma_A^{\prime T} U_i\sigma_B\quad \forall \quad \sigma_i^\prime \in \Delta_i\}.\end{equation}
Therefore, the set of Nash equilibria of game $\Gamma$ is given by
$$NE(\Gamma)=\{(\sigma_A,\sigma_B):\sigma_A\in R_A^\Gamma(\sigma_B)\text{ and } \sigma_B\in R_B^\Gamma(\sigma_A)\}.$$

An outcome-contingent contract between player $i$
and the game miner $G$ is a matrix $D_i$ that specifies a (possibly negative) transfer
from $i$ to $G$ for every outcome of $\Gamma$. We assume $D_i$ is finite. If players use strategies
$(\sigma_A,\sigma_B)$, then under contract $D_i$ player $i$ expects to
pay $\sigma_A^T D_i\sigma_B$ to G. Defining
$U_i^{D_i}\equiv U_i-D_i$, player $i$'s expected payoff is $\sigma_A^T
U_i^{D_i}\sigma_B$.  Therefore, we can view $D_i$ as a transformation
of $\Gamma$. We write the \textit{post-contract} game as
$\Gamma(D,i)\equiv\{\{A,B\},\{X_A,X_B\},\{U_i^{D_i},U_j\}\}$ for $i\neq j$, where the second argument of the post-contract game identifies the player with whom $G$ has contracted.  Note, that we simplify this notation by dropping this second argument whenever the context makes clear the identity of the player with whom $G$ has contracted.  We write the set
of possible contracts as $\mathcal{D} \equiv
\mathbb{R}^{|X_A|}\times\mathbb{R}^{|X_B|}$. The notation $D_0$ denotes the null contract, where all entries are zero.

\section{Maximal Mining}
\label{sec:maxmine}

Before introducing a formal
strategic setting for game mining in the next section, we first
explore the way that a player $A$ and the game miner can work together
to extract gains from $\Gamma$. This is the simplest game mining scenario. 

The first step is to define the \textit{aggregate payoffs} from a contract. These are the amounts that
the contracting parties can earn in equilibrium and divide among
themselves. Suppose $A$ and $G$ are the contracting parties and $D_A$
is their contract. Then the aggregate payoff that is apportioned
between $A$ and $G$ is given by the payoff that $A$ gets at a NE in
$\Gamma(D_A)$ \textit{before} $A$ pays to $G$ the amount specified in
$D_A$.

\begin{definition}\label{mutpayset}
The \textit{aggregate payoff set} for $A$ and $G$ from $D_A$ is:
$$M_A(D_A)\equiv\{\sigma_A^T U_A\sigma_B:(\sigma_A,\sigma_B)\in
NE(\Gamma(D_A))\}$$
\end{definition}

$A$'s payoffs are the only difference between $\Gamma$ and the post-contract game $\Gamma(D_A)$.  Here $D_A$ distorts $U_A$, thereby distorting $A$'s best response function. By varying $D_A$, we can give $A$ any best response function in $\Gamma(D_A)$.  This includes best response functions that designate dominant strategies as well as indifferences. Therefore, when $A$ exclusively contracts with the game miner, any profile such that $B$ chooses a best response can arise as an equilibrium of $\Gamma(D_A)$ for some $D_A$.  The same thing can be said for the situation where $A$ has the exclusive opportunity to make commitments on mixed strategies.  

We denote by $M^\ast_A(D_A)$ the maximum of the aggregate payoff set
from $D_A$.  The maximum over all aggregate payoff sets is
$\mathcal{M}_A \equiv \max_{D_A \in \mathcal{D}}\{M^\ast_A(D_A)\}$. It
is the maximum that $A$ and $G$ can possibly have to divide among
themselves in any NE of any game in which they sign a contract. We
refer to this quantity as the \textit{maxagg} (maximum aggregate
payoff). The maxagg is the subject of our first result.

\begin{proposition}\label{maxagg1}
\begin{equation}\label{maxaggeq1}\mathcal{M}_A=\max_{\sigma_A, \sigma_B}\{\sigma_A^T U_A\sigma_B:\sigma_B\in R_B^\Gamma(\sigma_A)\}. \end{equation}
\end{proposition}

\begin{proof}
From the definition of maxagg we have:
\begin{equation}\label{defmaxagg} \mathcal{M}_A=\max_{D_A}\{\max\{\sigma_A^T U_A\sigma_B:(\sigma_A,\sigma_B)\in NE(\Gamma(D_A))\}\}\end{equation}
Recall that the contract $D_A$ does not affect $B$'s payoffs $U_B$. This means that $NE(\Gamma(D_A))=\{\sigma\in \Delta:\sigma_A\in R_A^{\Gamma(D_A)}(\sigma_B)\quad\text{and}\quad \sigma_B\in R_B^\Gamma(\sigma_A)\}$. The trouble is to choose $D_A$ so that $A$'s best response correspondence meets $B$'s best response function at the maximizers that correspond to $\mathcal{M}_A$, $(\sigma_A,\sigma_B)$. This problem is solved trivially by choosing $D_A$ such that $A$ is indifferent among all strategy pairs. Then every action of $A$ is a best response to every action of $B$, including $\sigma_B$, which, by assumption is in $R_B^\Gamma(\sigma_A)$.  

So equation \ref{defmaxagg} becomes
$$\mathcal{M}_A=\max_{D_A}\{\max\{\sigma_A^T U_A\sigma_B:\sigma_B\in R_B^\Gamma(\sigma_A)\}\},$$
which is the same as equation \ref{maxaggeq1} because the set $\{\sigma_A^T U_A\sigma_B:\sigma_B\in R_B^\Gamma(\sigma_A)\}$ is independent of $D_A$.
\end{proof}

Proposition \ref{maxagg1} means that to find the maximum aggregate payoffs for $A$ and G, we simply
search $B$'s best response correspondences to all of $A$'s moves for
the one giving maximum payoff to $A$. This allows us to restrict our
analysis to the values of $U_A$ along $B$'s best response
correspondence. Note that certain outcomes in the aggregate payoff set correspond to profiles in which $A$ does not choose a best response to $B$'s strategy.  Only $B$ is being forced to make a best-response. So, maxagg is
nothing more than an upper bound on what is possible for $A$ and G
to obtain by making a contract. 

%In the real world, a game miner would be concerned with downside
%risk of any given contract. That is, the game miner would be reluctant to sign a contract
%$D_A$ if the game $\Gamma(D_A)$ has NE in which $G$ loses
%money. Now consider the Subgame Perfect Nash Equilibrium (SPE) concept applied
%to the extensive form game in which $G$ first decides whether to accept a given contract $D_A$, and
%then the associated underlying game $\Gamma(D_A)$ is played if $G$ accepts the
%contract. Under
%that equilibrium concept, when deciding whether to accept $D_A$,
%G knows what NE of $\Gamma(D_A)$ would be played if  $G$ accepts. Hence, under that concept, $G$ is only
%concerned with her payoff as prescribed by the strategies of $A$ and
%$B$ in some single associated NE of $\Gamma(D_A)$ (see analysis below of SPE of game mining).  In the
%real world though,  if $\Gamma(D_A)$
%contains multiple
%NE, the a game miner does \emph{not} know with
%certainty which NE of $\Gamma(D_A)$ would be played if $G$ accepted the contact $D_A$. Due to this, in the real world, a ``conservative'' game miner G
%might choose a contract that maximizes the
%\emph{minimum} aggregate payoff to be divided between $A$ and G, 
%to minimize how bad the situation for $G$ could be in a ``worst case" NE
%of $\Gamma(D_A)$. 

In some cases, a game miner might be concerned with maximizing the minimum of the aggregate payoff set as a way of guaranteeing a sufficient minimum payoff.  We write the minimum of the aggregate payoff set $M_A(D_A)$ as
$\underline{M}_A(D_A)$. Maximizing $\underline{M}_A(D_A)$ over all
contracts $D_A$, we get the maximum minimum aggregate payoff
$\underline{\mathcal{M}}_A$, called the \textit{maxminagg}:

$$\underline{\mathcal{M}}_A \equiv \max_{D_A} \min_{\sigma}\{\sigma_A^T U_A\sigma_B:\sigma\in NE(\Gamma(D_A))\}.$$

Trivially, $\underline{\mathcal{M}}_A\geq \underline{M}_A(D_0)$. Comparing maxminagg with
maxagg, we also know that $\underline{\mathcal{M}}_A\leq
\mathcal{M}_A$. And when there exists a contract $D_A$
such that $M_A(D_A)= \{\mathcal{M}_A\}$, we have that
$\underline{\mathcal{M}}_A=\mathcal{M}_A$. That is, if there exists a
contract $D_A$ such that the only NE of $\Gamma(D_A)$ yields the
maxagg to $A$ and G, then the maxagg and maxminagg are the same.

The next example continues from example \ref{ex:2} in the introduction, using maxagg and maxminagg to illustrate a distinction between the equilibrium outcomes in game mining and commitments to play pure strategies.  In this example, maxagg is associated with a NE strategy for $A$, $\sigma_A$, that is a mixed strategy with full support. However in this example, a commitment by $A$ to play (or not play) certain pure strategies will never allow $A$ and $G$ to achieve the maxagg. 

On the other hand, contracts that achieve the maxagg may also give rise to a
NE with an aggregate payoff lower than the maxagg or the NE aggregate payoff for the original game. So we may expect that a conservative miner would avoid such contracts that achieve the maxagg but also have bad equilibria. To address this, we show that there are
contracts that yield a unique mixed NE $\sigma_A$ for which the aggregate payoff is arbitrarily close to the maxagg. Because this NE is unique, the aggregate payoff is also the maxminagg that concerns the conservative game miner. Therefore, a conservative game miner would choose such a contract over one that involves a commitment to play (or not play) a specific strategy. 

\begin{example}\label{examplemsne}
Consider again the game $\Gamma$ presented in Ex.~\ref{ex:2} above:
$$\begin{array}{c|c|c|}
\multicolumn{1}{c}{} & \multicolumn{1}{c}{H} & \multicolumn{1}{c}{L} \\
\cline{2-3}
H & 1,.5 & 2,0 \\
\cline{2-3}
L & 0,0 & 1,1 \\
\cline{2-3}
\end{array}$$
where $A$ is the row player. Write $p \equiv \sigma_A(H)$. Then $B$'s best response correspondence is
\begin{equation*}
R_B^\Gamma(p)=\begin{cases}
H & \text{if $p>\frac{2}{3}$}\\
L & \text{if $p<\frac{2}{3}$}\\
q\in[0,1] & \text{if $p=\frac{2}{3}$.}
\end{cases}
\end{equation*}
If $A$ chooses $p<\frac{2}{3}$, then $B$ will choose $L$, and the payoff to $A$ will be $2p + 1(1- p) = 1 + p$. Likewise, if $A$ chooses $p>\frac{2}{3}$, then $B$ will choose $H$, and the payoff will be $p$. If $A$ chooses $p=\frac{2}{3}$ then $B$ chooses any combination of $H$ and $L$ yielding payoffs to $A$ between $\frac{2}{3}$ and $\frac{5}{3}$. Therefore, the maximum payoff for $A$ along $R_B^\Gamma$ is when $\sigma_A = (p,1-p)=(\frac{2}{3},\frac{1}{3})$ and $\sigma_B=(0,1)$:
$$\mathcal{M}_A=\sigma_A^T U_A \sigma_B = \left[\begin{array}{c}
\frac{2}{3}\\
\frac{1}{3}
\end{array}\right]^T
\left[\begin{array}{c c}
1 & 2\\
0 & 1
\end{array}\right]
\left[\begin{array}{c}
0\\
1
\end{array}\right] = \frac{5}{3}.$$
In other words, the maxagg payoff for $A$ and $G$ is achieved by a
mixed strategy. The problem is that there is no contract $D_A$ such
that $\Gamma(D_A)$ has a unique NE that achieves the maxagg payoff to $A$ and that results in a nonnegative payoff for $G$.
Moreover, for those $D_A$'s such there is a NE of $\Gamma(D_A)$ with
$A$'s payoff equaling the maxagg payoff, there are other NE with $A$'s payoff less
than $1$, which is already $A$'s payoff in every NE of the original game $\Gamma$. So it
would appear that $A$ has no incentive to form a contract with Game
Mining, Inc. However, there are contracts that produce a unique NE under
which the
aggregate payoff is arbitrarily close to the maxagg of $5/3$. An example of such a
contract that gets arbitrarily close to the maxagg is
\begin{equation*}
D_A = \left[\begin{array}{ c c }
1.5 & .5-\varepsilon \\
0 & -.5
\end{array}\right].
\end{equation*}
With this contract $A$'s payoffs are now given by
\begin{equation*}
U_A^{D_A} = \left[\begin{array}{c c}
1 & 2\\
0 & 1
\end{array}\right]-
\left[\begin{array}{ c c }
1.5 & .5-\varepsilon \\
0 & -.5
\end{array}\right] = 
\left[\begin{array}{ c c }
-.5 & 1.5+\varepsilon \\
0 & 1.5
\end{array}\right]
\end{equation*}
There is a unique NE for all $\varepsilon > 0$. As $\varepsilon$ approaches zero, that NE approaches $(2/3,L)$, so the maxminagg approaches $5/3$. As mentioned previously, this outcome is also possible if $A$ can commit to playing his NE mixed strategy.

%The Coaseian outcome occurs when players can bargain to achieve the outcome that maximizes total welfare. Here the Coaseian outcome is for $A$ to pay $B$ to play $L$ for a price between 0 and 1. This differs from the outcome under game mining. Therefore, game mining is not just a way to facilitate Coaseian outcomes when players cannot directly bargain.  Rather the presence of a game miner transforms the strategic 
%setting in a way that cooperation cannot. Therefore, in this example, the game miner provides a service that the players could not provide themselves. Note that this outcome could not be achieved by allowing commitments to play mixed strategies, since such commitments can not be enforced in one-shot games.
\end{example}

%The maxminagg is a reasonable concept, especially when the game miner
%cannot know for certain which of multiple NE will be played by the
%players. However, in this paper we will rely exclusively on the SPE
%concept applied to extensive form games where the contracts are chosen
%before the ensuing game is played. This concept requires that when
%choosing a contract the game miner knows which of multiple NE will be
%adopted by the players in the following, underlying game. Accordingly,
%SPE says that the miner can perfectly forecast which NE of the
%underlying game gets played. Whether or not such perfect forecasting
%is realistic -- and it arguably is not -- it is demanded by the SPE
%concept. Therefore, from now on we depart from the maxminagg concept,
%leaving for future work an analysis of game mining that incorporates
%maxminagg more fully.

The preceding results consider the aggregate payoffs that $A$ and $G$ can achieve by contracting. We next address the way in which $A$ and $G$ are able to divide those aggregate payoffs. The following result says that they are able to incorporate any division of the aggregate payoffs directly into the contract without affecting the best response correspondence of $A$ or $B$. (Whether $A$ and $G$ would accept such a division is a different issue.)

\begin{proposition}\label{sharemaxagg}
For any $a\in \mathbb{R}$ and $\sigma^\ast=(\sigma_A^\ast,\sigma_B^\ast)$ such that $\sigma_B^\ast\in R_B^\Gamma(\sigma_A^\ast)$, there exists a contract $D_A^\ast$ such that:
\begin{enumerate}
%	\item $\sigma_A^\ast\in R_A^{\Gamma(D_A^\ast)}(\sigma_B^\ast)$, and
	\item $(\sigma_A^\ast, \sigma_B^\ast) \in NE(\Gamma(D_A^\ast))$, and
	\item $\sigma_A^{\ast T} U_A^{D_A^\ast}\sigma_B^\ast=a$.
\end{enumerate}
\end{proposition}
\begin{proof}
By proposition \ref{maxagg1} there exists a contract $D_A$ such that
$(\sigma^\ast_A, \sigma^\ast_B) \in NE(\Gamma(D_A))$. Let
$\underline{1}$ stand for a matrix all of whose entries are $1$. So
$(\sigma_A^\ast)^T \underline{1} \sigma_B^\ast = 1$. Therefore there
is a scalar $x \in {\mathbb{R}}$ such that $$\left(\sigma_A^{\ast T}
\left(U_A-D_A-x\underline{1}\right)\sigma_B^\ast\right)\underline{1} =
a\underline{1}.$$ This gives us $D_A^\ast \equiv D_A+x\underline{1}$. Since
$U_A^{D_A^\ast}=U_A^{D_A}-x\underline{1}$, $(\sigma_A^\ast,
\sigma_B^\ast) \in
NE(\Gamma(D^\ast_A))$. 
\end{proof}

Proposition \ref{sharemaxagg} says that the aggregate payoffs that $A$ and
G get by mining are not affected by a restriction on the way in
which $A$ and $G$ divide those payoffs. So the way $\mathcal{M}_A$ is
divided between $A$ and $G$ in equilibrium will be determined by
strategic rather than technical considerations. This will be
convenient when we introduce a formal strategic environment in the
next section.

In some settings, there is no contract such that $A$ and $G$ can both benefit in any NE of $\Gamma(D_A)$. This is always the case when $B$ has a strictly dominant strategy.\footnote{Later we introduce the possibility that $A$ and $B$ both make contracts with $G$ in a sequential structure.  Here, there are contracts such that $A$ and $G$ both benefit in $\Gamma(D_A,D_B)$ even though $B$ has a strictly dominant strategy.} The intuition is that $A$'s contract with $G$ will never change $B$'s payoffs. Therefore, $B$ will always play his dominant strategy, no matter what the contract says. Therefore
there is nothing that a contract can do to help $A$. This intuition is formalized in the following result.

\begin{proposition}\label{strictdom}
If $B$ has a strictly dominant strategy, then there is no contract $D_A$ such that $A$ and $G$ both strictly benefit in any NE of $\Gamma(D_A)$.
\end{proposition}

\begin{proof}
Suppose $\tilde{x}_B$ is a strictly dominant strategy for player $B$. Then $\tilde{x}_B= R_B^{\Gamma(D_A)}(\sigma_A)=R_B^{\Gamma}(\sigma_A)$ for all $\sigma_A\in\Delta_A$. Hence the set of NE in $\Gamma$ is
$$\{\sigma^\ast\in \Delta :  \sigma_A^\ast=\arg\!\max_{\sigma_A}\sigma_A^T U_A\tilde{x}_B\text{ and } \sigma_B^\ast=\tilde{x}_B\}$$
and the set of NE in $\Gamma(D_A)$ is 
$$\{\tilde{\sigma}\in \Delta :  \tilde{\sigma}_A=\arg\!\max_{\sigma_A}\sigma_A^T U_A^{D_A}\tilde{x}_B \text{ and } \tilde{\sigma}_B=\tilde{x}_B\}.$$
If $G$ benefits by entering contract $D_A$, then $\tilde{\sigma}_A ^T D_A\tilde{x}_B> 0$. 

But this means that ${\tilde{\sigma}}^T_A U_A {\tilde{x}}_B > {\tilde{\sigma}}^T_A U^{D_A}_A {\tilde{x}}_B$. Since $\sigma^{\ast T}_A U_A {\tilde{x}}_B \ge \tilde{\sigma}_A^T U_A {\tilde{x}}_B$, by combining
we have $\sigma_A^{\ast T} U_A\tilde{x}_B>\tilde{\sigma}_A^T U_A^{D_A}\tilde{x}_B$ for all $\sigma^\ast,\tilde{\sigma}$. Accordingly,
$A$ will not benefit by signing $D_A$. \end{proof}

Proposition \ref{strictdom} puts a restriction on the set of games $\Gamma$ in which $A$ will benefit from the services of a game miner. However, as is shown in later sections, there are ample opportunities for $A$ and $G$ to benefit from contracts when player $B$ has
a strictly dominant strategy. In general, such a situation requires that $B$ also has an opportunity to make a payoff-contingent contract with the miner.

The next result establishes limits on game mining when one player has a weakly dominant strategy.

\begin{proposition}\label{weakdom1}
If $B$ has a weakly dominant strategy, then there is no contract $D_A$
such that both $A$ and $G$ strictly benefit in every NE of
$\Gamma(D_A)$ compared to not signing any contract. 
\end{proposition}

\begin{proof}
By contradiction. Suppose $x^\ast_B$ is weakly dominant and $D_A$ is a contract such that both $A$ and $G$ benefit in every NE of $\Gamma(D_A)$. There is a NE $(x_A^\ast,x^\ast_B)$ of $\Gamma$. For every $x_A\in X_A$ we have that 
$$x_A ^T U_A x^\ast_B\leq x_A^\ast U_Ax^\ast_B.$$
If $A$ is better off in every NE of $\Gamma(D_A)$, then for $x_A\in R_A^{\Gamma(D_A)}(x^\ast_B)$
$$x_A U_A^{D_A}x^\ast_B> x_A^\ast U_Ax^\ast_B$$
which implies 
$$x_A D_Ax^\ast_B<0$$
which in turn contradicts the fact that $G$ is better off.
\end{proof}

Proposition \ref{weakdom1} tells us that $A$ and $G$ cannot eliminate the
risk of loss in a NE of $\Gamma(D_A)$ if $B$ has a weakly dominant
strategy in $\Gamma$. If $A$ and $G$ are both conservative and require that they gain in every NE of $\Gamma(D_A)$, then no contract will be made between them.

\section{Game Mining Bargaining Structures}
\label{sec:structures}

\subsection{$G$ Accepts One Contract}
\label{subsec:onecontract}

Consider a situation in which players $A$ and $B$ encounter each other
in a simultaneous move game, $\Gamma=\langle\{A,B\}\{X_A,X_B\}, \{U_A,U_B\}\rangle$.  There is only one external party, G, that is willing to accept publicly observable outcome contingent contracts. Before $A$ and $B$ play $\Gamma$, they simultaneously offer contracts to G. These contracts are called $D_A$ and $D_B$ respectively. 

After observing $D_A$ and $D_B$, $G$ chooses either $D_A$, $D_B$ or $D_0$ (the null contract). Players $A$ and $B$ observe this contract and recognize its legally binding nature. $A$ and $B$ then engage in the simultaneous move game $\Gamma(D, i)$. $\Gamma(D, i)$ is the \textit{post-contract subgame}, where again we simplify the notation by dropping the second argument when the identity of the player with whom $G$ has contracted is clear.

Formally, this is an extensive form game with three stages: 
\begin{description}
    \item[Stage One:] Players $A$ and $B$ simultaneously offer contracts $D_A$ and $D_B$ to G.
	\item[Stage Two:] $G$ chooses $D_A$, $D_B$ or the null contract $D_0$.
	\item[Stage Three:] Players $A$ and $B$ play $\Gamma(D,i)$.
\end{description}

A strategy $S_i$ for $i=A,B$ in the extensive form game
is a pair $S_i\equiv (D_i,s_i)$. The first component, $D_i\in
\mathcal{D}$, is the offer that $i$ makes to $G$ in the first stage.
The second component is a function from the space of all
possible contracts and contractees, $ \mathcal{D}\times\{A,B,0\}$, to the space of probability
distributions over $i$'s action set $X_i$, i.e. $s_i:\mathcal{D}\times \{A,B,0\} \mapsto
\Delta_i$.  In other words, $s_i(D, j)$ gives $i$'s strategy (pure or mixed) in the post-contract subgame induced by $G$ choosing contract $D$ from player $j$.  Here we also simplify the notation by dropping the second argument when the identity of the player with whom $G$ has contracted is clear, e.g. $s_A(D_B) = s_A(D_B, B)$. The profile of strategies of player $A$ and player $B$ are
written as $S_{-G}$ where $s_{-G}=(s_A,s_B)$.

In stage two, $G$ selects an element of the choice set $\mathcal{I}_G\equiv\{0,A,B\}$. $S_G$ is the function that takes as input the history $\mathcal{D}_G=(D_A,D_B)$ and returns an element of $\mathcal{I}_G$ indicating which player's contract has been chosen.  We let $D_{S_G(\mathcal{D}_G)}=0$ indicate the game miner chooses not to contract with either player.  

Given a full strategy profile $(S_A, S_B, S_G)$, G's payoffs are
$$U_G(S_A,S_B,S_G) = s_A(D_{S_G(\mathcal{D}_G)})^T
D_{S_G(\mathcal{D}_G)} s_B(D_{S_G(\mathcal{D}_G)}).$$ 
$D_{S_G(\mathcal{D}_G)}$ is G's stage two choice given the stage one actions $(D_A,D_B)$, and $s_A(D_{S_G(\mathcal{D}_G)})$ is $A$'s stage three reaction to that choice. For $i=A,B$, the payoffs are 
$$U_i(S_A,S_B,S_G) = s_A(D_{S_G(\mathcal{D}_G)})^T U_i^{D_{S_G(\mathcal{D}_G)}} s_B(D_{S_G(\mathcal{D}_G)}),$$
if $G$ chooses $i$'s contract and
$$U_i(S_A,S_B,S_G) = s_A(D_{S_G(\mathcal{D}_G)})^T U_i s_B(D_{S_G(\mathcal{D}_G)})$$
otherwise. As shorthand, we represent this extensive form game as
$$\Gamma_G\equiv\langle \{A,B,G\}, \Gamma,
\{\mathcal{S}_i\}_{i=A}^B,\mathcal{S}_G,U_G \rangle.$$

\begin{definition}A \textit{subgame perfect equilibrium (SPE)} of $\Gamma_G$ is a strategy profile $S=(S_A,S_B,S_G)$ such that:
\begin{enumerate}
    \item $(s_A(D_{S_G(\mathcal{D}_G)}),s_B(D_{S_G(\mathcal{D}_G)}))\in NE(\Gamma(D_{S_G(\mathcal{D}_G)}))$ for all contracts $D\in \mathbb{R}^2\times \mathbb{R}^2$.
	\item $S_G$ is optimal given $s_{-G}$ for all pairs $(D_A,D_B)$, i.e.
\begin{align*}
U_G(S_A,S_B,S_G&)  \geq   U_G(S_A,S_B,S^\prime_G)
\end{align*}
for all $S_G^\prime$.
	\item $D_A$ is optimal given $S_G$, $s_A$ and $S_B$, i.e.
\begin{align*}U_i((D_A,s_A),S_B,S_G)  \geq U_i((D^\prime_A,s_A),S_B,S_G) 
\end{align*}
for all $D_A^\prime$ (\emph{mutatis mutandi} for $B$).
\end{enumerate}
\end{definition}

We turn our attention to finding the maximum amount that can be mined from $\Gamma$. To do so, we introduce a concept that is related to the aggregate payoff set from definition \ref{mutpayset}:

\begin{definition}\label{mutpayfunc}
The \textit{aggregate payoff function} for $A$ and $G$ from $D_A$ is:
$$m_A(D_A|s_{-G}) \equiv s_A(D_A)U_A^{D_A}s_B(D_A).$$
\end{definition}

The aggregate payoff function differs from the aggregate payoff set. Whereas the aggregate payoff set includes payoffs for all NE of $\Gamma(D_A)$, the aggregate payoff function simply returns the sum of $A$ and G's payoffs when $s_{-G}$ is played in $\Gamma(D_A)$. For example, if $s_{-G}$ selects a NE of the post-contract subgame $\Gamma(D_A)$, then the aggregate payoff function $m_A(D_A|s_{-G})$ selects one element from the aggregate payoff set $M_A(D_A)$. We denote by $\hat{D}_A$ a contract that maximizes $A$'s aggregate payoff function. 
%We denote by $\mathcal{D}^\ast_A$ the set of all such maximizers. 

In an SPE, $G$ will choose whichever contract yields her the highest payoff as determined by $(s_A,s_B)$. Given that, $i$'s should offer more to $G$ than $j$'s offers whenever $j$'s contract offers less than $m_i(\hat{D}_i|s_{-G})-s_A(D_j)^T U_i s_B(D_j)$.  The most that $i$ will ever be willing to offer $G$ is therefore determined by finding the contract of $j$ that results in the smallest payoff for $i$, called $\underline{D}_j$. Following this logic reveals that, loosely speaking, $G$ will contract with the player that has the greatest willingness to pay. In other words, there will not be an SPE in which $G$ accepts a contract from one player while the other has a greater willingness to pay. From the players' willingness to pay, we get the maximum SPE payment to $G$ in the following proposition.

\begin{proposition}\label{SPEmonPay}
The upper bound SPE payment to $G$ is
\begin{equation*}
\bar{U}_G=\max_i\{\mathcal{M}_i-\min_{\sigma_A,\sigma_B}\{\sigma_A^T U_i\sigma_B: \sigma_i\in R_i^\Gamma(\sigma_{-i})\}\}
\end{equation*}
\end{proposition}

\begin{proof}
First, let $\underline{D}_j(s_{-G})\equiv\arg\!\min_{D_j}s_A(D_j)^T U_is_B(D_j)$. That is, $\underline{D}_j$ is the contract that minimizes $i$'s payoff given $s_{-G}$. Also let 
$$\delta_i(D_i|s_{-G})\equiv m_i(D_i|s_{-G})-m_i(D_0|s_{-G})$$
be the change in $i$'s payoff by going from $\Gamma$ to $\Gamma(D_i)$. Similarly define
$$\delta_i(D_i,D_j|s_{-G})\equiv m_i(D_i|s_{-G})-m_i(D_j|s_{-G})$$
as the change in $i$'s payoff by going from $\Gamma(D_j)$ to $\Gamma(D_i)$. 

The proof follows from the strategic considerations of the players. Either (1) neither player pays G, or (2) player $i$ pays G. In the case of (2), $i$ will offer $G$ no more than necessary, which is the minimum increment above what $G$ would get by accepting $j$'s offer, $D_j$.  Player $i$ will only be willing to pay this amount if it is less than the amount that she gains by changing the game from $\Gamma(D_j)$ to $\Gamma(D_i)$, i.e., $\delta_i(D_i,D_j|s_{-G}).$ 
This difference is maximized by choosing $D_j$ to minimize $i$'s payoff in $\Gamma(D_j)$ and choosing $D_i$ to maximize $i$'s payoff in $\Gamma(D_i)$. Given $s_{-G}$, these arguments are $\underline{D}_j$ and $\hat{D}_i$ respectively. So we have that the maximum $i$ will pay in an SPE of $\Gamma_G$ given $s_{-G}$ is $\delta_i(\hat{D}_i,\underline{D}_j|s_{-G})$. 

Maximizing $i$'s payoff over all functions $s_{-G}$ and contracts $D_i$ we get the maxagg $\mathcal{M}_i$. Minimizing $i$'s payoff over all functions $s_{-G}$ we get $\min_{s_{-G}}\underline{D}_j(s_{-G})$ where the minimizer $s^i_{-G}=\arg\!\min_{s_{-G}}s_A(\underline{D}_j(s_{-G}))^T U_is_B(\underline{D}_j(s_{-G}))$ yields $\underline{D}_j(s^i_{-G})$. However, we know that since $s^i_{-G}$ is part of an SPE, that $s^i_{-G}(\underline{D}_j)$ is a NE of $\Gamma(\underline{D}_j)$. Therefore, $s^i_i(\underline{D}_j)\in R_i^\Gamma(s^i_j(\underline{D}_j))$. In other words, the only requirement in constructing $s^i_{-G}$ is that $i$ is always playing a best response. This is because $D_j$ can be such that $s^i_j(\underline{D}_j)$ is a best response to $i$. Therefore, 
$$s^i_A(\underline{D}_j)^T U_is^i_B(\underline{D}_j)=\min_{\sigma_i,\sigma_j}\{\sigma_A^T U_i\sigma_B: \sigma_i\in R_i^\Gamma(\sigma_j)\}.$$
Putting this together with $i$'s maxagg and choosing $i$ we get the result.
\end{proof}

Proposition \ref{SPEmonPay} gives an upper bound on the amount that the monopolist game miner can extract from a game between $A$ and $B$. This is achieved when player $i$ chooses the contract that minimizes $j$'s payoff given the strategies for the resulting post-contract game, $\underline{D}_i$.  At the same time, the amount offered to the game miner by the contract $\underline{D}_i$ is equal to the maximum amount that $j$ is willing to pay to change the game from $\Gamma(\underline{D}_i)$ to $\Gamma(\hat{D}_j)$.  The amount paid to the game miner in this situation is bounded by the players' payoff functions. So a monopolist game miner cannot, in this situation, extract arbitrary profits. 

The SPE concept here allows for some behavior that is unreasonable from a trembling hand perspective. For example, in order for $G$ to achieve her maximum payment, $i$ must choose contract $\underline{D}_i$. 
%\textbf{JB: ??? $\underline{D}_A$ is undefined
%here. DW: It's actually defined before the proposition as ``the contract of $j$ that results in the smallest payoff for $i$, called $\underline{D}_j$.'' I inserted that after you mentioned it last time.} 
However, $i$ might actually \textit{prefer} the outcome under $\hat{D}_j$ to the outcome under $\underline{D}_i$. That is, to support the equilibrium in which $j$ offers her maximum willingness to pay, $i$ offers quite a bit of money to $G$ for a deal she wants not to take effect.  $i$'s offer is only a best response to $j$'s slightly greater offer because $G$ will choose $j$'s contract, so that this unreasonable offer by $i$, $\underline{D}_i$, will never be accepted by G.  But, if $G$ trembled and chose $\underline{D}_i$, the outcome could be disastrous for $i$. In short, for some $\Gamma_G$, there exist SPE in which $G$ achieves her maximum payoff only if one of the players acts in a manner that seems unreasonable. 
%\textbf{JB: Can we prove that this equilibrium isn't
%trembling hand perfect? DW: Good question. Don't know. }

Instead of considering all possible equilibria, perhaps a more reasonable set of outcomes is one in which players only offer contracts $\hat{D}_i$ ($i=A,B$) that maximize the aggregate payoff function, $m_i(D_i | s_{-G})$. 
%\textbf{JB: I'm afraid I'm lost. Is this the same $\hat{D}_i$
%as defined above? DW: Yes, it is. } 
It is reasonable to restrict the analysis to this set of contracts because at least one aggregate payoff maximizer, $\hat{D}_i$, is a best response to every $D_j$ and $s_{-G}$.    Note that in some games, this restriction on players' choice of contracts can reduce G's maximum SPE payoff. 

\begin{center}
\begin{figure}[h]%
\includegraphics[scale=.5]{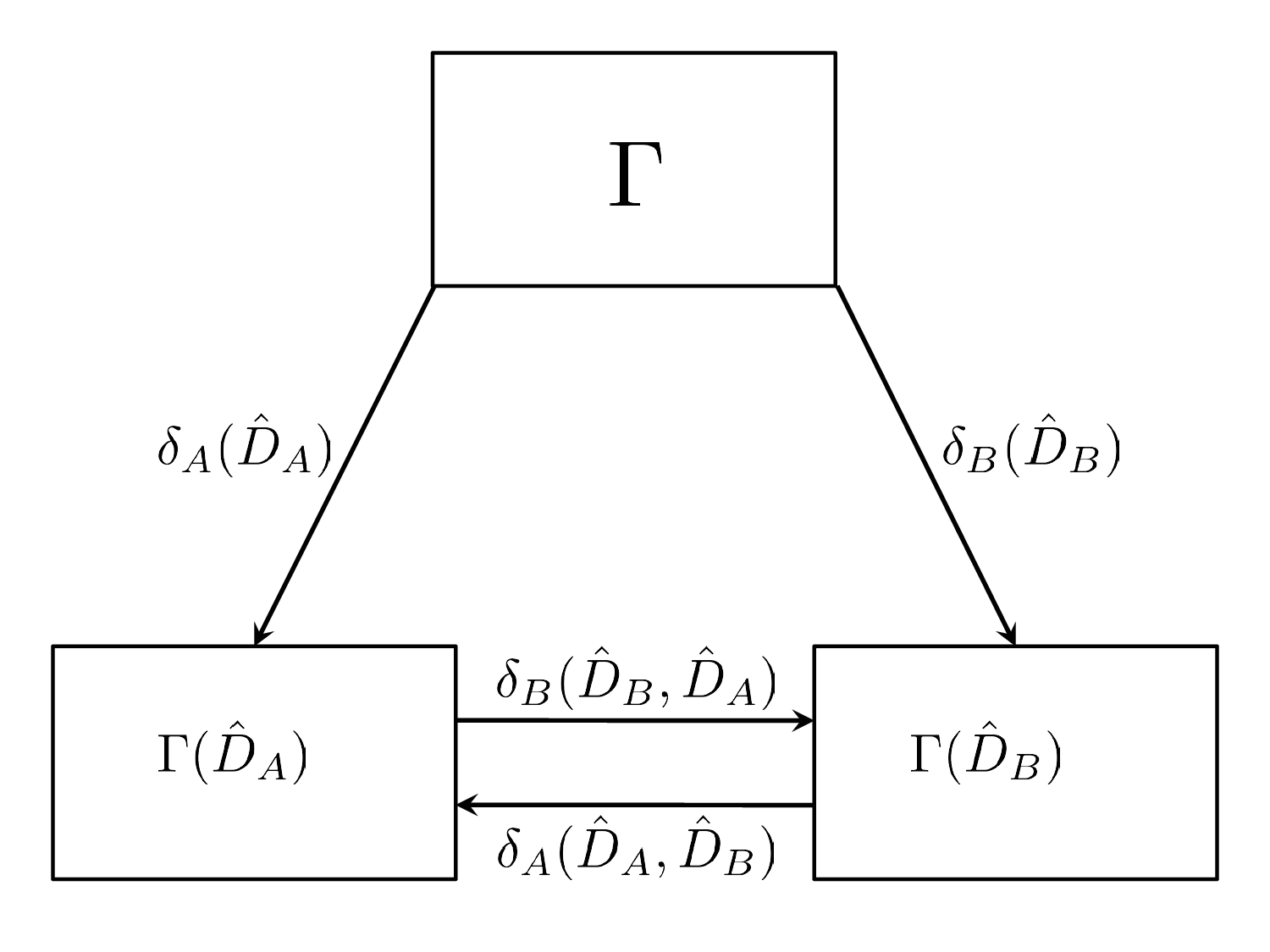}%
\caption{This diagram depicts a flow between $\Gamma$, $\Gamma(\hat{D}_A)$ and $\Gamma(\hat{D}_B)$ based on the willingness to pay of players $A$ and $B$.}
\label{fig:flowdiagram}
\end{figure}%
%\setlength{\unitlength}{3947sp}%
%\begingroup\makeatletter\ifx\SetFigFont\undefined%
%\gdef\SetFigFont#1#2#3#4#5{%
%  \reset@font\fontsize{#1}{#2pt}%
%  \fontfamily{#3}\fontseries{#4}\fontshape{#5}%
%  \selectfont}%
%\fi\endgroup%
%\begin{picture}(5562,3311)(2064,-2561)
%\put(6600,-2086){\makebox (0,0)[lb]{\smash{{\SetFigFont{12}{14.4}{\rmdefault}{\mddefault}{\updefault}{$\Gamma(\hat{D}_B)$}%
%}}}}
%\put(4350,-1800){\makebox (0,0)[lb]{\smash{{\SetFigFont{12}{14.4}{\rmdefault}{\mddefault}{\updefault}{$\delta_A(\hat{D}_A,\hat{D}_B)$}%
%}}}}
%\put(4350,-2400){\makebox (0,0)[lb]{\smash{{\SetFigFont{12}{14.4}{\rmdefault}{\mddefault}{\updefault}{$\delta_B(\hat{D}_B,\hat{D}_A)$}%
%}}}}
%\put(2514,-2049){\makebox (0,0)[lb]{\smash{{\SetFigFont{12}{14.4}{\rmdefault}{\mddefault}{\updefault}{$\Gamma(\hat{D}_A)$}%
%}}}}
%\put(4750,176){\makebox (0,0)[lb]{\smash{{\SetFigFont{12}{14.4}{\rmdefault}{\mddefault}{\updefault}{$\Gamma$}%
%}}}}
%\put(6100,-900){\makebox (0,0)[lb]{\smash{{\SetFigFont{12}{14.4}{\rmdefault}{\mddefault}{\updefault}{$\delta_B(\hat{D}_B)$}%
%}}}}
%\put(3000,-900){\makebox (0,0)[lb]{\smash{{\SetFigFont{12}{14.4}{\rmdefault}{\mddefault}{\updefault}{$\delta_A(\hat{D}_A)$}%
%}}}}
%\put(4920,-1300){\makebox (0,0)[lb]{\smash{{\SetFigFont{12}{14.4}{\rmdefault}{\mddefault}{\updefault}{$\delta_A(D_0,\hat{D}_B)$}%
%}}}}
%\put(4000,-1000){\makebox (0,0)[lb]{\smash{{\SetFigFont{12}{14.4}{\rmdefault}{\mddefault}{\updefault}{$\delta_B(D_0,\hat{D}_A)$}%
%}}}}
%\end{picture}%
\end{center}

Figure \ref{fig:flow} is a flow diagram that illustrates how the aggregate payoff maximizing contracts translate into the monopolist miner's payoffs.  Positive quantities represent movements in the direction of the associated arrow. So if $\delta_B(\hat{D}_B)>0$, we know that $B$ is willing to pay to change the game from $\Gamma$ to $\Gamma(\hat{D}_B)$. Hence, $\Gamma$ will not be the post-contract subgame in an SPE. Next, if $\delta_B(\hat{D}_B,\hat{D}_A)>\delta_A(\hat{D}_A,\hat{D}_B)>0$, then $B$ is willing to pay more to change the game from $\Gamma(\hat{D}_A)$ to $\Gamma(\hat{D}_B)$ than $A$ is willing to pay to change it from $\Gamma(\hat{D}_B)$ to $\Gamma(\hat{D}_A)$. $B$ will offer $G$ a contract such that G's payment is just greater than $A$ is willing to pay to change the game from $\Gamma(\hat{D}_B)$ to $\Gamma(\hat{D}_A)$. 

Another implication of proposition \ref{SPEmonPay} is that G's payoff can be greater than $\max\{\mathcal{M}_A,\mathcal{M}_B\}$. In other words, the winning contract may pay $G$ more than the maxagg for either player. The following example illustrates how a monopolist miner can make both players worse off than they were without the opportunity to mine. We demonstrate that this is the case even when players are restricted to choosing contracts that maximize aggregate payoff functions.

\begin{example}\label{fig:flow}
Consider the game $\Gamma$:
$$\begin{array}{c|c|c|c|}
\multicolumn{1}{c}{} & \multicolumn{1}{c}{x} & \multicolumn{1}{c}{y} & \multicolumn{1}{c}{z}\\
\cline{2-4}
x & -1,2 & -1,3 & 0,0 \\
\cline{2-4}
y & -1,-1 & 0,0 & 3,-1 \\
\cline{2-4}
z & -2,-2 & -1,-1 & 2,-1\\
\cline{2-4}
\end{array}$$
where $A$ is the row player. The unique NE of $\Gamma$ is $(y,y)$. Calculating $A$'s aggregate payoff function for $\hat{D}_A$, $\hat{D}_B$ and $\hat{D}_0$ as well as $A$'s willingness to pay, we get:
\begin{align*}
m_A(\hat{D}_A|s_{-G})=2, \quad m_A(\hat{D}_B|s_{-G})=-1 \quad\text{and}\quad m_A(D_0|s_{-G})=0\\
\Rightarrow \quad \delta_A(\hat{D}_A,\hat{D}_B)=3 \quad\text{and}\quad \delta_A(\hat{D}_A)=2
\end{align*}
By symmetry the quantities for $B$ are the same as the corresponding quantities of $A$.

The fact that $\delta_i(\hat{D}_i)=2>0$ for $i=A,B$ means that both players are willing to pay to change the game from $\Gamma$ to $\Gamma(\hat{D}_i)$, so $D_0$ will not be the outcome. Next, because $\delta_i(\hat{D}_i,\hat{D}_j)=3>0$ for $i=A,B$, we know that $G$ will get a payoff of $\delta_i(\hat{D}_i,\hat{D}_j)=3$ in equilibrium. This payoff is greater than $\mathcal{M}_i-m_A(D_0|s_{-G})=2$. In other words, if $i$'s contract is accepted, then the contract between $i$ and $G$ pays $G$ more than the increase in aggregate payoffs $\mathcal{M}_i-m_A(D_0|s_{-G})=2$. The reason is that $i$ is paying to avoid having $\Gamma(\hat{D}_j)$ become the equilibrium game. 

We also observe that $s_A(\hat{D}_i)^T U_i^{D_i} s_B(\hat{D}_i)= -1 < 0 = m_A(D_0|s_{-G})$. This says that $i$ gets less by having the equilibrium contract with $G$ than $i$ would get if neither player had the opportunity to offer contracts. For $j$, the player that does not win the equilibrium contract, the SPE payoff is also $-1$. Therefore, the both players are worse off with the opportunity to mine.
\end{example}

\subsection{G Accepts Both Contracts}
\label{subsec:bothcontracts}

We now relax the assumption that $G$ must choose between $D_A$ and $D_B$. After all, if $G$ is a true monopolist game miner and there are gains to be made by simultaneously contracting with both parties, then $G$ will certainly want to do this. 

The strategies in $\Gamma_G$ must be modified to accommodate this new possibility. First, a strategy $S_G$ for $G$ selects an element of G's choice set $\mathcal{D}_G$ after the history $(D_A,D_B)\in\mathcal{D}^2$. Since $G$ can now choose to accept both contracts if she wishes, the choice set $\mathcal{D}_G$ is now given by:
$$\mathcal{D}_G = \{(D_A,D_B),(D_A,D_0),(D_0,D_B),(D_0,D_0)\}.$$
This induces the game $\Gamma(D_i,D_j)$ where $D_i$ ($i=A,0$) is the contract between $A$ and $G$ and $D_j$ ($j=B,0$) is the contract between $B$ and G. Therefore, the game $\Gamma(D_i,D_j)$ is one in which $A$'s preferences are $U_A^{D_i}$ and $B$'s preferences are $U_B^{D_j}$. This means that the stage three strategy profile $s_{-G}=(s_A,s_B)$ is defined on $\mathcal{D}^2$ so that $s_i:(\mathcal{D})^2\mapsto \Delta_i$ ($i=A,B$). In other words, players select a strategy for every possible post-contract subgame of the form $\Gamma(D_i,D_j)$.

We refer to the stage three game that is played in equilibrium of $\Gamma_G$ as the \textit{equilibrium game}. If $G$ accepts only $D_A$, then the equilibrium game is $\Gamma(D_A)$. If $G$ accepts $D_A$ and $D_B$, then the equilibrium game is $\Gamma(D_A,D_B)$ and so on. The game $\Gamma(D_A,D_B)$ was not possible when $G$ could only accept a single contract. However, when $G$ can accept both contracts it is possible.  

This raises the issue of determining how $A$ chooses $D_A$ given that $B$ is choosing $D_B$. Given a function $s_{-G}=(s_A,s_B)$, $B$'s contract $D_B$ and G's decision $S_G$, $A$ chooses a contract in order to maximize his payoff. \begin{align}\label{AsNprob}
\max_{D_A} &\quad  s_A(D_{S_G(\mathcal{D}_G)})^T U_A^{D_A}s_B(D_{S_G(\mathcal{D}_G)})
\end{align}
This gives rise to a best response correspondence for $A$.
\begin{definition}
Player $A$'s \textit{best contract response correspondence given $s_{-G}$} is a set valued function $\Phi_A(\cdot |s_{-G}):\mathcal{D}\mapsto 2^\mathcal{D}$ that gives all of the contracts $D_A$ that maximize equation \ref{AsNprob} when $B$ makes contract $D_B$ given $s_{-G}$.
\end{definition}

By requiring that $s_{-G}$ selects a NE of every post-contract subgame, we guarantee that $s_A(D_{S_G(\mathcal{D}_G)})$ is a best response to $s_A(D_{S_G(\mathcal{D}_G)})$ and vice versa. When $s_{-G}$ meets this requirement, the best contract response correspondence amounts to a best response correspondence for the extensive form game. The following result uses the concept of a best contract response correspondence to categorize a monopolist game miner's payoffs when able to accept both contracts.

\begin{proposition}\label{SPEmonPayZ}
The monopolist game miner's equilibrium payoffs under the restriction that $G$ can only accept one contract are always as good and sometimes better than her payoffs without that restriction.
\end{proposition}

\begin{proof}
Suppose $\delta_A(\hat{D}_A)\geq 0$ and/or  $\delta_B(\hat{D}_B)\geq 0$ and $\delta_A(\hat{D}_A,\hat{D}_B)\geq \delta_B(\hat{D}_B,\hat{D}_A)\geq 0$, then with the restriction, $G$ gets $\delta_B(\hat{D}_B,\hat{D}_A)$. However, without the restriction, there is the possibility that for some $D_A$ and $D_B$, $A$ and $B$ both prefer $\Gamma(D_A,D_B)$ to $\Gamma(\hat{D}_A)$ and $\Gamma(\hat{D}_B)$. If $D_A\in \Phi_A(D_B)$ and $D_B\in \Phi_B(D_A)$ given $s_{-G}$, then this will be an equilibrium. When the equilibrium game is $\Gamma(D_A,D_B)$, neither player is paying for exclusivity, so G's payoff is zero instead of $\delta_B(D_B,D_A)$.

Further, the threat of an outcome $\Gamma(D_A,D_B)$ can never induce $i$ to pay more than $\delta_j(\hat{D}_j,\hat{D}_i)$ for exclusivity. This is because $\delta_j(\hat{D}_j,\hat{D}_i)$ is the value for $j$ of going from $\Gamma(\hat{D}_A)$ to $\Gamma(\hat{D}_B)$. Given that $i$ pays for exclusivity, there is no payment that $j$ can make to change the game from $\Gamma(\hat{D}_A)$ to $\Gamma(D_A,D_B)$ because $i$'s contract is contingent on exclusivity.
\end{proof}

Proposition \ref{SPEmonPayZ} says that a monopolist game miner cannot be made worse off by restricting herself to accept a single contract. The reason is that when $G$ does not restrict herself, then she does not give up $D_A$ in order to accept $D_B$. Therefore, if $G$ accepts $D_B$, then her best response is to accept any contract $D_A$ for which her payoff under $\Gamma(D_A,D_B)$ is at least her payoff under $\Gamma(D_B)$. Knowing this, $A$ will choose $D_A$ such that G's payoff under $\Gamma(D_A,D_B)$ is exactly what it is under $\Gamma(D_B)$. The same holds for $B$. Therefore, $G$ is made worse-off by the ability to make contracts with both players. Put differently, the threat of an equilibrium game $\Gamma(D_A,D_B)$ never induces the players to pay more, and it is sometimes better for the players. 

The above suggests that the one-contract restriction might be the result of payoff maximizing behavior. That is, G's payoff in equilibrium of the one-contract game might be equivalent to a payment not to contract with the other player. Hence, the restricted game is equivalent to a game in which $A$ and $B$ submit two-element stage-one offers, $(D_i,z_i)$, where $D_i$ is the matrix of strategy-contingent transfers and $z_i$ is a payment not to make a contract with $j$. If $z_i=0$, then $i$ places no exclusivity restriction on G's acceptance of $D_i$. Therefore, G's payoff from accepting $A$'s contract is $z_A+s_A(D_A,D_0)^T D_A s_B(D_A,D_0)$. If $z_A=z_B=0$ then G's payoff from accepting both contracts is $s_A(D_A,D_B)^T (D_A+D_B)s_B(D_A,D_B)$.

\subsection{Sequential Contracts}
\label{subsec:sequential}

We now examine the role of timing on game mining outcomes. The game is exactly as previously described, except that $A$ first selects a contract to be observed by $B$ before $B$ selects a contract. In this setting we find that $A$ may have a first-mover advantage and also that contracts are not equivalent to the pre-commitments of \cite{Renou09}. Both points are demonstrated in the following example.

\begin{example}
Consider the game $\Gamma$
\begin{table}[h]
\centering
\begin{tabular}{c | c | c | c |}
\multicolumn{1}{c}{} & \multicolumn{1}{c}{x} & \multicolumn{1}{c}{y} & \multicolumn{1}{c}{z} \\
\cline{2-4}
x & 2,5 & 0,0 & 5,4 \\ \cline{2-4}
y & 1,3 & 1,2 & 2,0 \\ \cline{2-4}
z & 0,3 & 0,1 & 2,0 \\ \cline{2-4}
\end{tabular}
\end{table}
where $A$ is the row player. The unique NE of this game is $(x,x)$. Note that $x$ is a strictly dominant strategy for $B$. By proposition \ref{strictdom}, there is no contract $D_A$ such that $A$ gets a better payoff in a NE of $\Gamma(D_A)$ than in a NE of $\Gamma$. Despite this fact, there is a contract $D_A$ such that 
$$s_A(D_A,D^\ast_B)^T U_A^{D_A}s_B (D_A,D^\ast_B)> s_A(D_0)^T U_As_B(D_0)$$ 
where $D^\ast_B \in \Phi_B(D_A \mid s_{-G})$ is a best contract response to $D_A$ given $s_{-G}$. In other words, there is a contract $D_A$ such that when $B$ chooses his best contract response to $D_A$, $A$ gets a higher payoff in $\Gamma(D_A,D^\ast_B)$ than in any NE of $\Gamma$. To illustrate, suppose $A$ signs a contract with $G$ to pay $G$ $2$ whenever the outcome is $(x,x)$, then the unique NE of $\Gamma(D_A)$ is $(y,x)$. The post contract game, $\Gamma(D_A)$, is given by:
\begin{table}[h]
\centering
\begin{tabular}{c | c | c | c |}
\multicolumn{1}{c}{} & \multicolumn{1}{c}{x} & \multicolumn{1}{c}{y} & \multicolumn{1}{c}{z} \\
\cline{2-4}
x & 0,5 & 0,0 & 5,4 \\ \cline{2-4}
y & 1,3 & 1,2 & 2,0 \\ \cline{2-4}
z & 0,3 & 0,1 & 2,0 \\ \cline{2-4}
\end{tabular}
\end{table}

Then $B$'s best contract response is a contract $D^\ast_B$ that promises to pay $G$ $4$ if the outcome is $(y,x)$ and $3$ if the outcome is $(y,y)$. This will make $(x,z)$ the unique NE of $\Gamma(D_A,D^\ast_B)$. 

\begin{table}[h]
\centering
\begin{tabular}{c | c | c | c |}
\multicolumn{1}{c}{} & \multicolumn{1}{c}{x} & \multicolumn{1}{c}{y} & \multicolumn{1}{c}{z} \\
\cline{2-4}
x & 0,5 & 0,0 & 5,4 \\ \cline{2-4}
y & 1,-1 & 1,-1 & 2,0 \\ \cline{2-4}
z & 0,3 & 0,1 & 2,0 \\ \cline{2-4}
\end{tabular}
\end{table}

The final outcome gives $A$ his highest payoff from $U_A$, and it gives $B$ his maxagg. $G$ gets zero in the equilibrium of $\Gamma$, $\Gamma(D_A$, and $\Gamma(D_A,D^\ast_B)$. Therefore, this outcome is the unique SPE of $\Gamma_G$.
\end{example}

Note that if $B$ was the first to select a contract, then $B$ would choose $D_0$ to which $A$'s best contract response is $D_0$. Both players would be willing to pay for the right to move first. $A$ would pay the difference between his payoff in $\Gamma(D_A, D^\ast_B(D_A \mid s_{-G})$ and $\Gamma$, which is $5$.  However $B$ would only be willing to pay the difference between his payoff in $\Gamma$ and $\Gamma(D_A, D^\ast_B)$, which is just $1$.

This example draws a sharp distinction between game mining and pre-commitments to play or not play certain strategies. Suppose $A$ instead selected a contract that made $x$ a never-best-response. Then $B$'s best response is $D_0$, and the outcome is $(y,x)$, which is worse for $A$. $A$ does not want to commit to not playing $x$ because $(x,z)$ is the ultimate goal. He rather wants to commit to $(x,x)$ not being the outcome, so that $B$ will commit to $(y,x)$ and $(y,y)$ not being the outcome. 

Exploiting contract timing is yet another way that game miners game miners can extract profits from players even when players are making the offers. Since $A$ has a first-mover advantage, and $B$ has a second-mover disadvantage, both are willing to pay to move first. Suppose $A$ recognizes this advantage before $B$ and approaches $G$ with his desired contract $D_A$. $G$ could potentially put $A$ on hold and notify $B$ to start a bidding war over the first-mover advantage. The first-mover advantage is worth more to $A$ than it is to $B$, five versus one, so $A$ would end up paying $B$'s maximum willingness to pay for the first-mover advantage. In contrast to the discussion above where players make simultaneous offers and $G$ accepts one contract, this type of bidding war can happen even though $G$ is free to accept both contracts.

\subsection{G Makes Offers}
\label{subsec:Gmakes}

Until this point we have assumed a particular bargaining structure in which $A$ and $B$ make take-it-or-leave-it offers to G. This implies that G's only bargaining power is in rejecting contracts that result in negative payoffs. Suppose now that we change the game so that $G$ makes publicly observable offers to $A$ and $B$. Then $A$ and $B$ simultaneously accept or reject the offers $G$ has made. So $A$ and $B$ will now accept any contract that does not make them worse off, given the other's choice. This clearly places more power in the hands of G. 

To accommodate the new structure of the game, we alter the definition of strategies. Now G's stage one strategy is $s_G\in \mathcal{D}^2$. $A$ and $B$ have binary stage two strategies $s_i^2:\mathcal{D}^2\mapsto \{\text{accept},\text{reject}\}$ and stage-three strategies $s_i^3:\mathcal{D}^2\mapsto \Delta_i$ ($i=A,B$), which we sometimes shorten to be $s_{-G}^2$ and $s_{-G}^3$. So $G$ selects a contract for each player, $s_G$. Then each player chooses to accept or reject the contract they are offered, $s_{-G}^2$. Finally, the players play the post-contract game according to $s_{-G}^3$.

We want to characterize G's payoffs in an SPE. To do so, consider the following devious plan where $G$ can sometimes create a high-order prisoner's dilemma between $A$ and $B$. This is illustrated in the example below.

\begin{example}
Consider the game $\Gamma$ where $A$ is the row player and both players have a strictly dominant strategy to choose $z$.
$$\begin{array}{c|c|c|c|}
\multicolumn{1}{c}{} & \multicolumn{1}{c}{x} & \multicolumn{1}{c}{y} & \multicolumn{1}{c}{z} \\
\cline{2-4}
x & 5,5 & 1,5 & 1,6 \\
\cline{2-4}
y & 5,1 & 1,1 & 2,2 \\
\cline{2-4}
z & 6,1 & 2,2 & 3,3 \\
\cline{2-4}
\end{array}.$$ 
The unique NE is $(z,z)$, and both players get a payoff of $3$. Then if $G$ offers the following contract to $A$
$$D_A=\left[\begin{array}{c c c}
2.99 & -2 & 0 \\
3 & -1 & -3 \\
4 & .5 & 0 \\
\end{array}\right]
$$
the unique NE of $\Gamma(D_A)$ is $(y,z)$. $A$'s payoff in $\Gamma(D_A)$ is $2-(-3)=5$, so $A$ would accept $D_A$, getting $5$ rather than $3$. $B$'s payoff in $\Gamma(D_A)$ is $2$. If $G$ offers $D_B = D_A^T$ to $B$, then the equilibrium of $\Gamma(D_A,D_B)$ is $(x,x)$. The payoff to $B$ is $5-2.99 = 2.01$. Therefore, $B$ would accept $D_B$ given that $A$ accepts $D_A$ because he will get $2.01$ rather than $2$. By the symmetry of $\Gamma$, $D_A$ and $D_B$, we know that each player $i$ prefers to accept $D_i$ regardless of whether $j\neq i$ accepts or rejects his offered contract.

This is very similar to a prisoner's dilemma game because each player $i=A,B$ has an incentive to accept $D_i$ regardless of whether $j\neq i$ accepts $D_j$. This moves the players from $\Gamma$, where the unique NE gives them $(3,3)$, to $\Gamma(D_A,D_B)$, where the only NE gives them $(2.01,2.01)$. By playing $A$ against $B$ the game miner gets $2(5-2.01)=5.98$.\footnote{Of course, the game miner could reduce the contract payoffs for $D_i(w,w)$ ($i=A,B$) from 2.01 to $2+\epsilon$, where $\epsilon>0$ is an arbitrarily small number, and the game miner's payoff would strictly increase without changing any of the equilibria. The number 2.01 is used here for simplicity.} The situation can be visualized alternatively as the following PD game where $A$ is the row player and $B$ is the column player, and the payoffs are given by the unique NE of the resulting post-contract games:

$$\begin{array}{c|c|c|}
\multicolumn{1}{c}{} & \multicolumn{1}{c}{\text{accept }D_B} & \multicolumn{1}{c}{\text{reject }D_B} \\
\cline{2-3}
\text{accept }D_A & 2.01,2.01 & 5,2 \\
\cline{2-3}
\text{reject } D_A& 2,5 & 3,3 \\
\cline{2-3}
\end{array}.$$ 

Note that $G$ once again moves the game from an inefficient equilibrium to an efficient one. However, as a profit maximizer, the game miner captures the efficiency gains.
\end{example}

The example above shows that $G$ can potentially do better for herself by selecting contracts that both $A$ and $B$ will accept than by contracting with one player exclusively. The intuition for why this is possible is that $G$ relies on the fact that $\Gamma(D_A)$ and $\Gamma(D_B)$ will never obtain in equilibrium. Therefore, $G$ is free to offer contracts $D_A$ and $D_B$ such that she loses money in the unique NE of $\Gamma(D_A)$ and $\Gamma(D_B)$. This allows her the flexibility to make sure the NE of $\Gamma(D_A,D_B)$ is in her favor. Note that this result does not rely on the fact that both players have strictly dominant strategies. It is simple to construct examples for games in which only one player has a dominant strategy or neither player has a dominant strategy.

The following proposition provides an upper bound on the game miner's payoff when she extracts profits according to this scheme.

\begin{proposition}
The maximum that a monopolist game miner can profit by offering contracts $(D_A,D_B)$ such that $A$ and $B$ have weakly dominant strategies to accept is
$$\max_{x_A,x_B} x_A^T(U_A+U_B)x_B-\min_{x^\prime_A,x^\prime_B}\{x^{\prime T}_A U_B x^\prime_B: x_B^\prime\in R_B^\Gamma(x_A^\prime)\}-\min_{x_A^{\prime\prime},x_B^{\prime\prime}} \{x_A^{\prime\prime T} U_Bx_B^{\prime\prime}: x_A^{\prime\prime}\in R_A^\Gamma(x_B^{\prime\prime})\}$$
\end{proposition}

\begin{proof}
The first term is the maximum amount that $A$ and $B$ can earn in any outcome of $\Gamma$. The second term is the minimum amount that $G$ can force $B$ to get by designing $D_A$ such that $x_A^\prime$ is a best response to $x_B^\prime$. This is  because $G$ is constrained so that $x_B^\prime$ must lie on $B$'s best response correspondence for $\Gamma$. The third term is the equivalent of the second term for $A$ rather than $B$. 

Suppose $G$ earns more than this maximum, then
\begin{align}\label{Gsprofiteqn}
s_A(D_A,D_B)^T (D_A+D_B)s_B(D_A,D_B)>\max_{x_A,x_B} x_A^T(U_A+U_B)x_B...\\
-\min_{x^\prime_A,x^\prime_B}\{x^{\prime T}_A U_B x^\prime_B: x_B^\prime\in R_B^\Gamma(x_A^\prime)\}-\min_{x_A^{\prime\prime},x_B^{\prime\prime}} \{x_A^{\prime\prime T} U_Bx_B^{\prime\prime}: x_A^{\prime\prime}\in R_A^\Gamma(x_B^{\prime\prime})\}\notag
\end{align}
but because $A$ and $B$ each have a weakly dominant strategy to accept, we know that
$$s_A(D_A,D_B)^T (U_A-D_A)s_B(D_A,D_B)\geq s_A(D_0,D_B)^T U_A s_B(D_0,D_B)$$
and
$$s_A(D_A,D_B)^T (U_B-D_B)s_B(D_A,D_B)\geq s_A(D_A,D_0)^T U_B s_B(D_A,D_0).$$
%We can rewrite these as
%$$s_A(D_A,D_B)^T U_As_B(D_A,D_B)-s_A(D_0,D_B)^T U_A s_B(D_0,D_B)\geq s_A(D_A,D_B)^T D_As_B(D_A,D_B)$$
%and
%$$s_A(D_A,D_B)^T U_Bs_B(D_A,D_B)-s_A(D_A,D_0)^T U_B s_B(D_A,D_0)\geq s_A(D_A,D_B)^T D_Bs_B(D_A,D_B)$$
Together, these inequalities imply
\begin{align*}
s_A(D_A,D_B) ^T (D_A+D_B)s_B(D_A,D_B)&\leq s_A(D_A,D_B)^T (U_A+U_B) s_B(D_A,D_B)...\\
-s_A(D_0,D_B)&^T U_A s_B(D_0,D_B)- s_A(D_A,D_0)^T U_B s_B(D_A,D_0).
\end{align*}
The left-hand-side is G's profit, and the maximum of the right-hand-side is given by the right-hand-side of inequality \ref{Gsprofiteqn}. Therefore, we have a contradiction.
\end{proof}

This result is important because it says that the game miner cannot make an arbitrary profit from the players by giving each a dominant strategy to accept her offer. Therefore, the game miner's profit and any efficiency loss for the players are always limited.

\section{Discussion}
\label{sec:discussion}

As mentioned in the introduction, the game mining analysis opens up several new research areas. One natural extension is to consider game mining situations in which there are more than two players or even with multiple game miners. With multiple players simultaneously choosing contracts, the blocking and exclusivity concerns we address above are likely to become much more complicated. An open question is how the market structure affects the game miner's opportunity to profit.\footnote{An earlier version of this paper analyzes perfect competition and duopoly competition among game miners}

%Another research area concerns the difficulty of modeling the game miner's uncertainty over which outcome will obtain in the game $\Gamma(D_A)$ after signing the contract $D_A$ with player $A$. In an SPE analysis, the game miner's decision to sign or not sign $D_A$ occurs through the process of backward induction. That is, the SPE approach assumes the game miner somehow knows the outcome of $\Gamma(D_A)$, $s_{-G}(D_A)$, before she signs the contract with $A$. In the real world, the game miner would likely not be so certain about future events. In fact, the game miner's beliefs would likely assign nonzero probability to the occurrence of non-equilibrium outcomes of $\Gamma(D_A)$. Hence, rather than an equilibrium-approach, it may be valuable to adopt a statistical approach to game mining, such as the Predictive Game Theory (PGT) models described in \cite{Wolpert08}.

Yet another research question is whether replacing the structured bargaining between players and game miners with unstructured bargaining will change the profitability of game mining. One might also consider the reverse situation in which the stage game $\Gamma$ is a game of unstructured bargaining between $A$ and $B$, but the negotiations between players and game miners follow structured bargaining. Here, we would have that players sign contracts through structured bargaining in an attempt to gain an advantage in the unstructured bargaining that follows. The question is how would players design contracts to distort their utility possibilities set in such a way that benefits them in the ensuing unstructured bargaining. An interesting technical issue arises in deciding whether to allow players to sign contracts that lead to utility possibility sets that are nonconvex. If so, then a solution concept other than the Nash Bargaining Solution is needed [see \cite{Nash50a,Kalai75}].

All of this raises a crucial question: Why aren't real game mining firms wreaking havoc on real markets? Game mining appears to be very possible according to basic game theory, so if it is not generally possible in the real world, what assumptions are being violated?

There are many potential answers to this question. One tempting explanation is that the payoff structure of most real world games makes them unable to be mined. This seems a strange assertion because, as shown, even games in which both players have strictly dominant strategies can be mined for profit depending on the market structure.  Other potential answers are that the calculations are too difficult in practice, that the time frame in real world games is too short, that game mining could be considered illegal, that imperfect information limits game mining opportunity, or that some kind of strategic uncertainty makes game mining impractical. These explanations should be explored in future work because they might shed light on the way game theory applies to real world strategic settings. 

There are other questions to explore. For instance, does game mining imply that certain games should never exist because the minute they appear they will be mined into an alternate game? In some sense this gives rise to a meta-game whereby a player that finds himself involved in an easily mineable game might assume that the game will be mined and therefore conclude that he is actually playing a different game. Or, in a game with multiple equilibria, one equilibrium might make the game susceptible to mining by an outside party that ultimately makes both parties worse off (like what happens when a game miner makes offers that give players strictly dominant strategies to accept). Therefore, that susceptible equilibrium might become less likely than an equilibrium that is more robust. In this way game mining introduces an equilibrium refinement: choose the equilibrium that makes game mining least profitable.

These questions and others are not only interesting for their ability to shed light on game mining concepts, but also more generally for their ability to shed light on the noncooperative theory.

%\begin{acknowledgements}
%We would like to thank Mark Wilber and Nicholas Shunda for helpful discussion.
%\end{acknowledgements}

\bibliographystyle{spbasic}      % basic style, author-year citations
\bibliography{/Users/jamesbono/Desktop/ThesisBib}   % name your BibTeX data base

%\begin{thebibliography}{}
% etc
%\end{thebibliography}

\end{document}